\documentclass[review,hidelinks,onefignum,onetabnum]{siamart220329}
\nolinenumbers

\usepackage{lipsum}
\usepackage{amsfonts}
\usepackage{graphicx}
\usepackage{epstopdf}
\usepackage{algorithmic}
\ifpdf
  \DeclareGraphicsExtensions{.eps,.pdf,.png,.jpg}
\else
  \DeclareGraphicsExtensions{.eps}
\fi

\newsiamremark{remark}{Remark}
\newsiamremark{hypothesis}{Hypothesis}
\crefname{hypothesis}{Hypothesis}{Hypotheses}
\newsiamthm{claim}{Claim}

\headers{Epps Effect and the Signature of Short-Term Momentum Traders}{J. Busca and L. Thomir}

\title{Epps Effect and the Signature of Short-Term Momentum Traders\thanks{Submitted to the editors 9/11/2023.}}

\author{Jérôme Busca\thanks{JB Quantitative Solutions, LLC 
  (\email{jerome.busca@protonmail.com})}
\and Léon Thomir\thanks{CentraleSupélec, work completed while a student at ESILV 
  (\email{thomir.leon@gmail.com})}}

\usepackage{amsopn}

\ifpdf
\hypersetup{
  pdftitle={Epps Effect and the Signature of Short-Term Momentum Traders},
  pdfauthor={J. Busca and L. Thomir}
}
\fi

\begin{document}

\maketitle

\begin{abstract}
 It is a well-documented fact that the correlation function of the returns on two "related" assets is generally increasing as a function of the horizon $h$ of these returns. This phenomenon, termed the Epps Effect, holds true in a wide variety of markets, and there is a large body of literature devoted to its theoretical justification. Our focus here is to describe and understand a deviation to the Epps effect, observed in the context of the foreign exchange and cryptocurrency markets. Specifically, we document a sharp local maximum of the cross-correlation function of returns on the Euro EUR/USD and Bitcoin BTC/USD pairs as a function of $h$. Our claim is that this anomaly reveals the activity of short-term momentum traders. 
\end{abstract}



\section{Introduction}
	\label{sec:intro}

    In 1979, Epps \cite{Epps} highlighted the presence of a significant drop in the correlation between stocks when decreasing the time horizon of returns $h$. This phenomenon was first observed on stocks \cite{Zebedee,Chan,Bonanno_2001} and then in other markets such as foreign exchange \cite{Lundin}. Several theoretical justifications have been proposed to account for the phenomenon. The Epps effect can, for instance, be explained by a lead-lag phenomenon among specific stocks \cite{Toth_2006}, or by the asynchronous nature of ticks in liquid markets \cite{Reno}, although Tóth et al. established in a subsequent paper that tick asynchrony can't fully explains the formation of Epps curves \cite{Epps_revisited} (an observation that was confirmed in a recent study \cite{2021_Chang}).
     
    Our focus in this paper is the cross-correlation function of the returns on the EUR/USD (Euro rate in US dollar) and BTC/USD (Bitcoin price) pairs. This choice was driven by the fact that each pair is the flagship asset in its own market --- forex and crypto-currencies, respectively --- and that one can strongly suspect an interesting interplay between the traditional currency market, and the relatively new crypto-currency market. 
    
    It is widely known that the collective actions of traders play a central role in the dynamics of financial markets. The interactions and decisions of individual traders, motivated by a myriad of factors, combine to form a composite force known as the market factor. We will use this key insight to model traders' actions to first order as buying and selling both assets EUR/USD and BTC/USD at the same time, in the manner of an index (this can also be seen as "buying or selling the dollar").
    
     This paper is designed as follows. In \cref{sec:analysis}, we introduce the data set we use and compute the experimental cross-correlation function $\rho(h)$. In \cref{sec:model}, we use a simple Gaussian model to explain the peak we observe in $\rho$. Finally, in \cref{sec:simulation}, we build a more realistic agent-based Monte Carlo simulation which, once calibrated, shows good agreement with the data.

    \section{Empirical Analysis}  
    \label{sec:analysis}
    In this section, we conduct an empirical analysis on forex exchange and cryptocurrency markets through the leading pairs EUR/USD and BTC/USD and highlight the presence of the Epps curve, along with a deviation to it. 
    
    \subsection{Data}

    To be specific, we actually chose to conduct our analysis on the EUR/USDT and BTC/USDT pairs. Tether (USDT) is a type of cryptocurrency referred to as a stablecoin, designed to have a value which is pegged to the US Dollar. Our choice was guided by the availability of high-quality data sets, as well as the necessity to avoid data issues such as asynchrony between the FX and crypto markets, among other factors.
    
    We used a best bid/best offer data set from Binance (the largest crypto exchange), provided by Tardis, over the period 27 November 2020 -- 19 July 2022. As the cryptocurrency market operates continuously, unlike the FX market, we removed weekends to align Bitcoin data with FX data. Then, we calculated the simple-mid prices $P(t)$ for further analysis.
    
    \subsection{Correlations}

    We define the log-return of length $h$ for an asset of mid price $P(t)$ to be:
    \begin{align*}
    r_h(t)= \log\left(\frac{P(t)}{P(t-h)} \right),
    \end{align*} 
    
    and we denote by $\rho(h)$ the correlations between the log-returns on EUR/USDT (1) and BTC/USDT (2) as a function of the length $h$: \\
    \begin{equation}\label{rho_empirical}
    \rho(h) = \frac{\left\langle \left(r_{h}^{1}-\langle r_{h}^{1}\rangle\right)\left(r_{h}^{2}-\langle r_{h}^{2}\rangle\right)\right\rangle}{ \sigma^{1}\sigma^{2} },
    \end{equation}
    \\
    where the bracket $\langle . \rangle$ denotes time average, and the standard deviations of the returns are defined as usual by:
    \begin{align*}
    \sigma^k=\sqrt{\left\langle \left(r^k_h\right)^2 \right\rangle - {\langle r^k_h \rangle}^2},\ \ \ k=1,2.
    \end{align*} 

    We also compute a confidence interval at the 95\% level around (\ref{rho_empirical}) using a standard Fisher transformation \cite{Fisher}.
    
    \subsection{Empirical Results} 
    
    \begin{figure}[hbt!]
    \centering
    \includegraphics[width=0.5\textwidth]{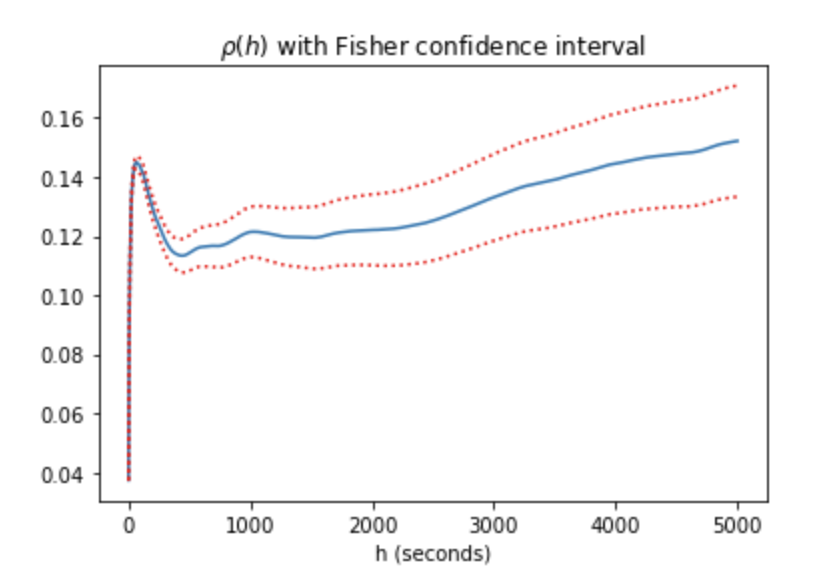}
    \caption{Correlation between EUR/USDT and BTC/USDT returns as a function of horizon h (\ref{rho_empirical})}
    \label{fig:rho_empirical_full_figure}
    \end{figure}

\begin{figure}[H]
    \centering
    \includegraphics[width=0.5\textwidth]{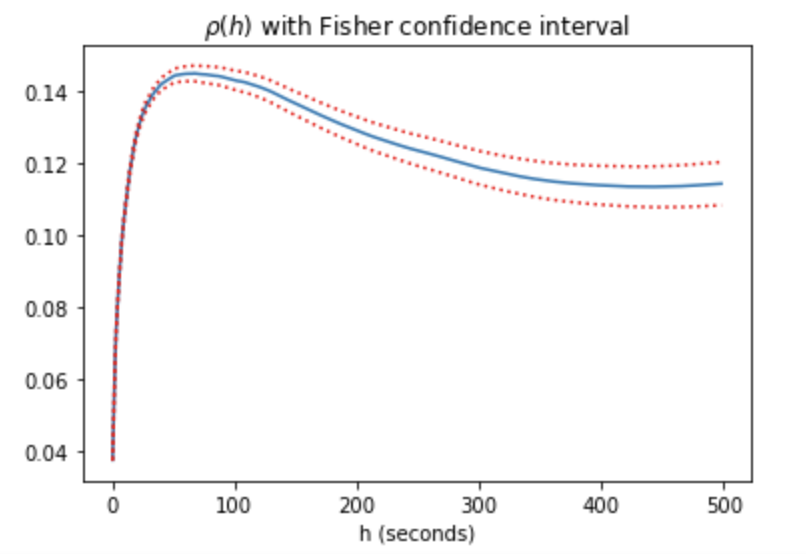}
    \caption{Correlation between EUR/USDT and BTC/USDT returns as a function of horizon h (\ref{rho_empirical})}
    \label{fig:rho_empirical_figure}
    \end{figure}

    Figure~\ref{fig:rho_empirical_full_figure} represents the cross-correlation $\rho(h)$ between EUR/USDT and BTC/USDT returns. We observe a classic Epps effect --- an overall increase in correlation with $h$. However, if we zoom in on higher frequencies (Figure~\ref{fig:rho_empirical_figure}), we observe an anomaly which manifests itself as a sharp peak around 60 seconds, with value 0.15, followed by a decrease to 0.12. These fluctuations are statistically significant.
    
    \section{A Gaussian Model of Momentum} 
    \label{sec:model}

    In this section we describe a simple Gaussian model which explains how momentum traders' activity can generate a peak in the cross-correlation function. We assume in the following there are only two assets, with prices $s_t=(s^1_t,s^2_t)$, and that the market consists of: i) a momentum trader; ii) a noise trader; and iii) a market-maker. The momentum trader is assumed to trade solely in an equally-weighted index based on $s^1$ and $s^2$ and to use a simple momentum rule with window $\tau>0$. For computational ease, we assume the trader looks at the simple returns (as opposed to percentage or log returns) of the two assets. With these rules, his inventory (in both asset 1 and 2) at time $t$ is given by:
    \begin{equation}\label{MM_inventory}
    p_t = \bar{p}\left(s^1_t - s^1_{t-\tau} + s^2_t - s^2_{t-\tau}\right),
    \end{equation}    
    where $\bar{p}>0$ is a constant. As for the noise trader, we model his inventory $X_t$ using an Ornstein-Uhlenbeck process with zero mean:
    \begin{equation}\label{OU}
    dX^k_t = -\lambda X^k_t dt + \sigma dW^k_t,\ \ k=1,2,
    \end{equation}
    with $\lambda,\sigma>0$ constant (independent of $k$ for simplicity) and $W^1_t$, $W^2_t$ independent Brownians, and with independent Gaussian initial conditions  equal to the stationary distribution :
    \begin{equation}\label{OU_initial_condition} 
    X^k_0\,{\buildrel d \over=}\,{\cal N}\left(0,\sigma^2/2\lambda\right),\ \  k=1,2. 
    \end{equation}
The assumption that the noise trader's inventory should follow (\ref{OU}) is fairly natural and can be shown to hold, for instance, in the classic Avellaneda-Stoikov model \cite{Avellaneda}, in the limit of large speed of trading (when Poisson processes reach their diffusion limit).
    
    Lastly, the market-maker is simply the counterparty (liquidity provider) to the momentum and noise traders. If we further assume a linear market impact for all liquidity-taking trades, with elasticity $\theta>0$, along with some noise with volatility $\nu>0$, we can easily write down the price process increment for the two assets:
    \begin{equation}\label{price_process}
    ds^k_t = \theta\left(dp_t + dX^k_t\right) + \nu dZ^k_t,\ \ k=1,2,
    \end{equation}
    where $Z^k_t$ are standard Brownian motions, independent of each other and of the $W^k_t$'s. Integrating (\ref{price_process}) and choosing initial conditions that don't generate extra constants for simplicity, we get: 
    \begin{equation}\label{state_equation}
    \left\{
    \begin{split}
    p_t &= \bar{p}\left(s^1_t - s^1_{t-\tau} + s^2_t - s^2_{t-\tau}\right) \\ 
    s^k_t &= \theta\left(p_t + X^k_t\right) + \nu Z^k_t,\ \ k=1,2.
    \end{split}
    \right.
    \end{equation}

    The price process therefore satisfies:
    \begin{equation}\label{price_process_final}
    s^k_t = \varepsilon\left(s^1_t - s^1_{t-\tau} + s^2_t - s^2_{t-\tau}\right) + \theta X^k_t + \nu Z^k_t, \ \ k=1,2,
    \end{equation}
    where $\varepsilon = \bar{p}\theta>0$ can be interpreted as a non-dimensional coupling parameter of the system. In the following, we always assume $\varepsilon\ll 1$ and expand all relevant quantities to first order in $\varepsilon$. From (\ref{price_process_final}), the price process can be rewritten
    \begin{equation}\label{price_process_epsilon}
    \left\{
    \begin{split}
    (1-\varepsilon) s^1_t - \varepsilon s^2_t &= -\varepsilon\left(s^1_{t-\tau} + s^2_{t-\tau}\right) + \theta X^1_t + \nu Z^1_t\\
    \\
    -\varepsilon s^1_t + (1-\varepsilon) s^2_t &= -\varepsilon\left(s^1_{t-\tau} + s^2_{t-\tau}\right) + \theta X^2_t + \nu Z^2_t.
    \end{split}
    \right.
    \end{equation}
    Noting that $\frac{1-\varepsilon}{1-2\varepsilon} \simeq 1+\varepsilon$, from (\ref{price_process_epsilon}) we have, to first order in $\varepsilon$
    \begin{equation}\label{approx_price_process}
    \left\{
    \begin{split}
    s^1_t &\simeq -\varepsilon \left(s^1_{t-\tau} + s^2_{t-\tau}\right) + \left(1+\varepsilon\right)\left(\theta X^1_t + \nu Z^1_t\right) + \varepsilon\left(\theta X^2_t + \nu Z^2_t\right)\\
    \\
    s^2_t &\simeq -\varepsilon \left(s^1_{t-\tau} + s^2_{t-\tau}\right) + \varepsilon\left(\theta X^1_t + \nu Z^1_t\right) + \left(1 + \varepsilon\right)\left(\theta X^2_t + \nu Z^2_t\right),\\
    \end{split}
    \right.
    \end{equation}
    so that
    \begin{equation}\label{price_process_explicit}
    \begin{split}
    s^k_t &\simeq \theta X^k_t + \nu Z^k_t + \varepsilon\theta \left(X^1_t - X^1_{t-\tau}\right) + \varepsilon\nu\left(Z^1_t - Z^1_{t-\tau}\right) 
    \\&+ \varepsilon\theta\left(X^2_t - X^2_{t-\tau}\right) + \varepsilon\nu\left(Z^2_t - Z^2_{t-\tau}\right),\ \ k=1,2 \\
    \end{split}
    \end{equation}
    and the $h$-horizon return is given by:
    \begin{equation}\label{h_return}
    \begin{split}
    s^k_{t+h} - s^k_t &= \theta\left(X^k_{t+h} - X^k_t\right) + \nu\left(Z^k_{t+h} - Z^k_t\right)\\
    &+\varepsilon\theta\left(\left(X^1_{t+h} - X^1_t\right) - \left(X^1_{t+h-\tau} - X^1_{t-\tau}\right)\right)\\ 
    &+\varepsilon\theta\left(\left(X^2_{t+h} - X^2_t\right) - \left(X^2_{t+h-\tau} - X^2_{t-\tau}\right)\right)\\
    &+\varepsilon\nu\left(\left(Z^1_{t+h} - Z^1_t\right) - \left(Z^1_{t+h-\tau} - Z^1_{t-\tau}\right)\right)\\
    &+\varepsilon\nu\left(\left(Z^2_{t+h} - Z^2_t\right) - \left(Z^2_{t+h-\tau} - Z^2_{t-\tau}\right)\right)\\
    \end{split}
    \end{equation}
    The quantities of interest are
    \begin{equation}\label{c_kl_definition}
    c_{kl}(h) := \langle s^k_{t+h} - s^k_t, s^l_{t+h} - s^l_t\rangle,\ \ k,l = 1,2,
    \end{equation}
    as well as the cross-correlation function 
    \begin{equation}\label{rho_definition}
    \rho(h) := \frac{c_{12}(h)}{\sqrt{c_{11}(h) c_{22}(h)}}
    \end{equation}
    
    \noindent We can now state our main result, an explicit formula for the cross-correlation function (\ref{rho_definition}). 

    \ 
    
    \noindent{\bf Main Result}
    \begin{theorem}\label{main_result}
    To first order in $\varepsilon$, we have: 
    \begin{equation}\label{rho_main_result}
    \frac{1}{2\varepsilon} \rho(h) = \frac{\theta^2\left(1-e^{-\lambda h}-e^{-\lambda\tau} + \frac{1}{2}e^{-\lambda(h+\tau)}+\frac{1}{2}e^{-\lambda\left|h-\tau\right|}\right) + \xi h\wedge\tau}{\theta^2\left(1-e^{-\lambda h}\right) + \xi h},
    \end{equation}
    where $h\wedge \tau = \mathrm{min}(h,\tau)$, and $\xi$ is the parameter $\xi = \frac{\nu^2}{\sigma^2/\lambda}$
    \end{theorem}

    To establish this result, we will need the following lemma.

    \begin{lemma}\label{main_lemma}
    If $m\in\left\{0,1\right\}$, we have, for all $t,h>0$:
    \begin{equation}\label{Z} 
    \langle Z_{t+h} - Z_t, Z_{t+h-m\tau} - Z_{t-m\tau}\rangle = \left(h - m\tau\right)_+,
    \end{equation}
    where $a_+ = \mathrm{max}(a,0)$; and
    \begin{equation}\label{X}
    \langle X_{t+h} - X_t, X_{t+h-m\tau} - X_{t-m\tau}\rangle = \frac{\sigma^2}{2\lambda}\left(2e^{-\lambda m\tau} - e^{-\lambda(h+m\tau)} - e^{-\lambda\left|h - m\tau\right|}\right),
    \end{equation}
    where $Z_t$ is either $Z^1_t$ or $Z^2_t$, and $X_t$ is either $X^1_t$ or $X^2_t$.  
    \end{lemma}

    \noindent{\bf Proof of Lemma \ref{main_lemma}}

    \noindent $Z$ is a standard Brownian motion. Thus, using ordinary stochastic calculus
    \begin{equation*}
    \langle Z_t,Z_s\rangle = t\wedge s \hbox{\ \ for all } s,t>0.
    \end{equation*}
    Therefore
    \begin{equation*}
    \begin{split}
    \langle Z_{t+h} - Z_t, Z_{t+h-m\tau} - Z_{t-m\tau}\rangle &= (t+h-m\tau) + (t-m\tau) \\
    &- t\wedge(t+h-m\tau)-(t-m\tau)\wedge(t+h) \\ 
    &= (t+h-m\tau) + (t-m\tau) + (-t)\vee (-t-h+m\tau) \\
    &+ (-t+m\tau)\vee(-t-h),
    \end{split}
    \end{equation*}
    where $a\vee b=\hbox{max}(a,b)$. Since $\mu + a\vee b = (a+\mu)\vee(b+\mu)$ for all $\mu$, we find the above is equal to
    \begin{equation*}
    (h-m\tau)_+ + (-h-m\tau)_+ = (h-m\tau)_+,
    \end{equation*}
    since $m,h,\tau\ge 0$, which establishes (\ref{Z}). 

    \noindent $X$ is an Ornstein-Uhlenbeck process following (\ref{OU}) and (\ref{OU_initial_condition}). Therefore
    \begin{equation*} 
    X_t = e^{-\lambda t} X_0 + \sigma\int_0^te^{-\lambda(t-u)} dW_u,
    \end{equation*}
    and for all $s,t>0$ we have
    \begin{equation*}
    \begin{split}
    \langle X_t, X_s\rangle &= e^{-\lambda(t+s)}\frac{\sigma^2}{2\lambda} + \sigma^2 e^{-\lambda(t+s)}\int_0^{t\wedge s} e^{2\lambda u}du\\
    &= e^{-\lambda(t+s)}\frac{\sigma^2}{2\lambda}\left(1 + \left(e^{2\lambda t\wedge s} - 1\right)\right)\\
    &= \frac{\sigma^2}{2\lambda} e^{-\lambda\left|t-s\right|},
    \end{split}
    \end{equation*}
    since $t+s-2t\wedge s = \left|t-s\right|$, from which (\ref{X}) follows easily. 

    \ 

    Let's now prove the main result, Theorem \ref{main_result}. By symmetry $c_{11}(h) = c_{22}(h)$ for all $h>0$ and, using (\ref{h_return}) and Lemma \ref{main_lemma}, we have, to first order in $\varepsilon$:
    \begin{equation}\label{c_diagonal}
    c_{11}(h) = c_{22}(h) = \frac{\sigma^2\theta^2}{\lambda}\left(1 - e^{-\lambda h}\right) + \nu^2 h\\
    \end{equation}
    and
    \begin{equation}\label{c_off_diagonal}
    \frac{1}{2\varepsilon} c_{12}(h) = \frac{\sigma^2\theta^2}{\lambda}\left(1-e^{-\lambda h}-e^{-\lambda\tau} + \frac{1}{2} e^{-\lambda(h+\tau)}+\frac{1}{2}e^{-\lambda\left|h-\tau\right|}\right) +\nu^2\left(h - (h-\tau)_+\right);
    \end{equation}
    and since $h - (h-\tau)_+ = h\wedge\tau$, combining (\ref{c_diagonal}) and (\ref{c_off_diagonal}) and dividing by $\sigma^2/\lambda$, we proved (\ref{rho_main_result}).

    \ 

    Now that we have an explicit formula for the cross-correlation function $\rho(h)$ with (\ref{rho_main_result}), we can plot its typical shape. The plot below illustrates $\rho(h)$ for the following values of the parameters: $\lambda=0.03162$, $\xi=0.0001$, $\theta=0.6$, $\tau=66$, $\varepsilon=0.0505$. These values come from the calibration of the model to our data set (see next section).

    \begin{figure}[hbt!]
    \centering
    \includegraphics[width=0.5\textwidth]{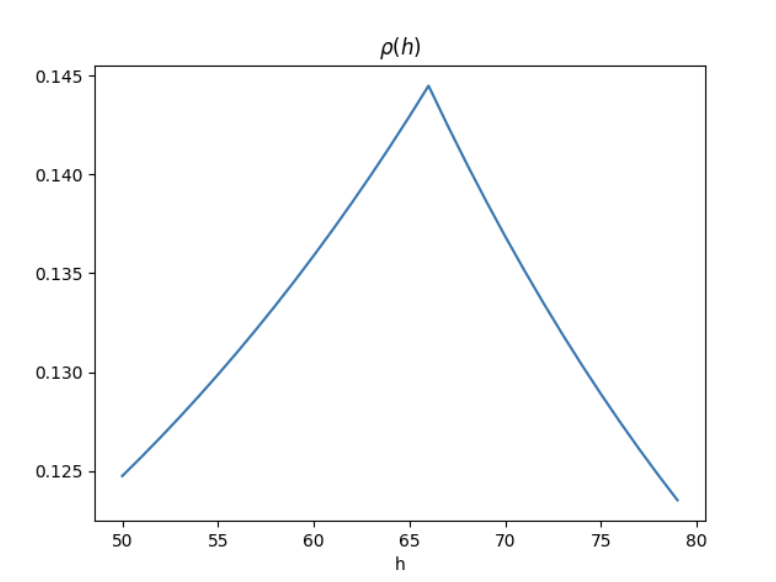}
    \caption{Cross-correlation $\rho(h)$ in (\ref{rho_main_result})}
    \label{fig:Gaussian_rho}
    \end{figure}

    We observe on Figure~\ref{fig:Gaussian_rho} that the cross-correlation $\rho(h)$ in (\ref{rho_main_result}) {\bf exhibits a sharp peak (kink) at } $\mathbf{h=\tau}$, {\bf the trading horizon of the momentum trader}. We therefore established --- in the simplified setting of our Gaussian model --- how the signature of index momentum traders can easily be 
    detected with the help of the cross-correlation function.

    \section{Agent-based Simulation}
    \label{sec:simulation}

    In this section, we build an agent-based Monte Carlo simulation which extends our simplistic Gaussian model by incorporating more realistic features. As in the Gaussian model, 
    we assume in the following there are only two assets, with prices $s_t=(s^1_t,s^2_t)$, and that the market consists of: i) a momentum trader; ii) a noise trader; and iii) a market-maker.
    
    In this simulation, we relax the constraint that Poisson processes reach their diffusion limit. Thus, we use the "original" form of Avellaneda Stoikov's model \cite{Avellaneda}. \\
    On the other hand, to avoid unbounded inventories, we impose a cap on the momentum trader's positions. For the sake of realism, we also take into account the tick sizes $\eta$.

    \subsection{Noise trader}
    We model the noise trader following \cite{Avellaneda}. Specifically, we assume trade executions occur at the ask (resp. bid) price at a rate described by  Poisson processes $N_{t}^{a}$ and $N_{t}^{b}$  with respective intensity parameters $\lambda^{a,b}_t(\delta^{a,b}_t) = A^{a,b}e^{-k^{a,b}\delta^{a,b}_t}$, where $\delta^{a,b}_t$ is the half-spread between the mid and the ask (resp. bid) price, and $A^{a,b},k^{a,b}>0$ are constants measuring the liquidity of the market.

    The inventory $q_{t}^{n}$ is then simply given by
    \begin{align*}
                q_{t}^{n} = \psi^n (N_{t}^{a} - N_{t}^{b}),
    \end{align*}
    where $\psi^n>0$ is a parameter (trade size).

    \subsection{Momentum trader}

    We assume the momentum trader buys and sells an equally-weighted index based on $s^1$ and $s^2$, denoted by $\mathrm{index}_t$, and uses a simple momentum rule with window $\tau>0$. To manage risk, the momentum trader has a maximum position constraint. We also force a long (resp. short) position when the index moves back above (resp. below) the moving average.

    Their inventory $q_{t}^{m}$ is modeled as 
    \begin{equation*}
    q_{t}^{m} = \left\{
    \begin{split}
    &\mathrm{min}(\mathrm{max}(q_{t-1}^{m}, 0) + \psi^m,\ q_{\mathrm{max}}^m) \ \ \ \ \hbox{if} \ \ \mathrm{index}_{t-1} > \mu_{t-1}
    \\
    \\
    &\mathrm{max}(\mathrm{min}(q_{t-1}^{m}, 0) - \psi^m,\ -q_{\mathrm{max}}^m)  \ \ \hbox{otherwise,}
    \end{split}
    \right.    
    \end{equation*} 
    where $\mu_t$ is the $\tau-$moving average of $\mathrm{index}_t$, $\psi^m>0$ is the trade size, and $q_{\mathrm{max}}^m>0$, is the maximum absolute inventory.
  
    \subsection{Market maker}
    The market-maker is the liquidity provider. It is therefore modeled as the counterparty to the noise and momentum traders. His inventory 
    is simply $q_{t}^{mm} = - q_{t}^{n} - q_{t}^{m}$. He adjusts the half-spread $\delta^{a,b}_t$ as a function of k, q, the tick size $\eta$, the volatility, and a risk aversion parameter $\gamma$ as described in \cite{Avellaneda}. 

    Furthermore, the impact of liquidity-taking orders on the mid price is assumed to be linear with elasticity $\theta > 0$, on top of a Brownian noise with volatility $\nu>0$. Before rounding to tick size $\eta$, the mid price processes are therefore given by \\
    \begin{align*} 
        s_{t}^k = s_{t-1}^k + \theta^k (q_{t}^n + q_{t}^m - q_{t-1}^n - q_{t-1}^m) + \nu^k dZ^k_t, \ \ k=1,2,
    \end{align*}
    where $dZ^k_t$ are Brownian increments. 
    
    \subsection{Monte Carlo Simulation}
    We ran a discrete Monte-Carlo simulation of the model above, with time resolution $\mathrm{dt}=0.5s$, over a duration $T = 2 \cdot 10^7 \mathrm{dt}$, i.e. around 115 trading days. We took realistic initial values for $s^1$ and $s^2$, respectively $1.10$ USD and $30,000$ USD.
    
    We then calibrated the parameters of our model to our data set. Here are the fitted values: 
 $\eta=[1e-4, 1e-2]$, $A=[1, 1]$, $\theta=[2.7e-11, 2.7e-6]$, $\nu=[3e-5, 2.04]$, $\gamma=[6.46e-6, 3.47e-9], k=[3466, 34.66]$, $\psi^n=[100 000, 4]$, $\psi^m=[3 000 000, 120]$, $\tau=500s$, $q_{\mathrm{rm}}^{m}=[6 500 000, 250]$. 

    \

\begin{figure}[H]
    \centering
    \includegraphics[width=0.5\textwidth]{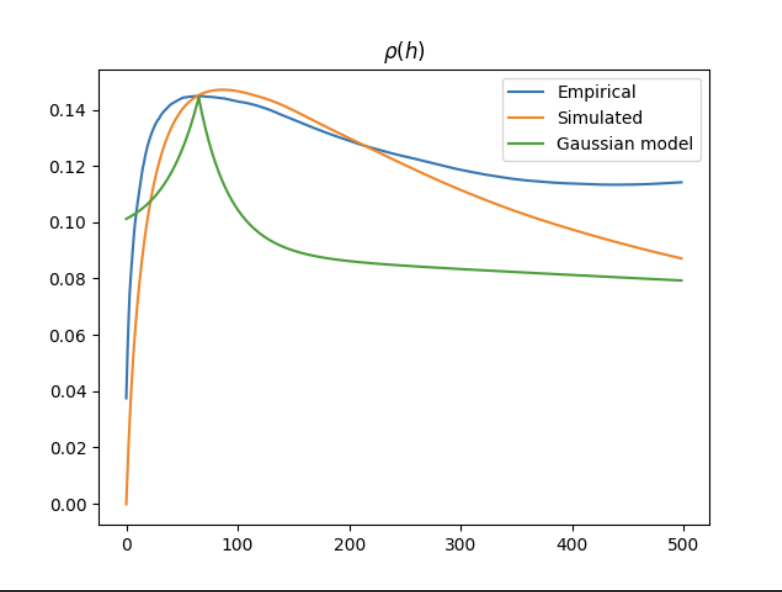}
    \caption{Comparison of the cross-correlation functions $\rho(h)$}
    \label{fig:all_correlations}
    \end{figure}

    Figure~\ref{fig:all_correlations} shows the empirical correlation function (blue), the agent-based simulated correlation function (orange) and the correlation function of our simple Gaussian model (green) as a function of horizon h, expressed in seconds.

    We observe that the cross-correlation function of our agent-based simulation exhibits a good fit to the data up to h = 120 seconds, before dropping and failing to maintain a correlation level of 0.12. We suspect that the impact of agents with longer trading horizons is the cause of a persistent correlation for larger values of $h$.

    As to the Gaussian model, we see a sharp peak around h = $\tau=66$ seconds, followed by a slow drop to the 0.08 level. Due to its limitations, it is impossible to create a "smoother" or larger peak in correlation.

    The main difference in the choice of parameters between the Gaussian and agent-based models is the value $h=\tau$, where the peak of $\rho$ is reached in the Gaussian model. Indeed, for the simulated model, it is necessary to set a higher value for $\tau$ because of the switch to a discrete model and the resulting effect on the noise trader's orders. We made available the Python code for the simulation at 
    
    \texttt{https://github.com/RimohtL/EppsEffect}

\bibliographystyle{siamplain}
\bibliography{references}
\end{document}